\documentstyle[twoside,fleqn,espcrc2,epsfig]{article}

\newcommand{\GF}{G_{\mbox{\rm {\tiny F}}}}
\newcommand{\h}{\mbox{\bf h}}
\newcommand{\hc}{\mbox{\rm h.c.}}

\newcommand{\lesssim}{ {\
\lower-1.2pt\vbox{\hbox{\rlap{$<$}\lower5pt\vbox{\hbox{$\sim$}}}}\ } }
\newcommand{\gtrsim}{ {\
\lower-1.2pt\vbox{\hbox{\rlap{$>$}\lower5pt\vbox{\hbox{$\sim$}}}}\ } }
\newcommand{\be}{\begin{equation}}
\newcommand{\ee}{\end{equation}}
\newcommand{\bea}{\begin{eqnarray}}
\newcommand{\eea}{\end{eqnarray}}

\newcommand{\nn}{\nonumber}

\newcommand{\cL}{{\cal L}}
\newcommand{\cO}{{\cal O}}
\newcommand{\cH}{{\cal H}}

\newcommand{\cA}{{\cal A}}

\newcommand{\tr}{\mbox{\rm tr}}

\newcommand{\GeV}{\mbox{\rm GeV}}

\newcommand{\al}{\alpha}
\newcommand{\als}{\alpha_{\mbox{\rm {\scriptsize s}}}}

\newcommand{\NC}{\mbox{\rm {\tiny NC}}}

\newcommand{\QCD}{QCD$(\infty)\;$}

\newcommand{\eff}{\mbox{\rm{\tiny eff}}}

\newcommand{\lbl}{\label}

\newcommand{\ra}{\rightarrow}

\newcommand{\AmS}{{\protect\the\textfont2
  A\kern-.1667em\lower.5ex\hbox{M}\kern-.125emS}}

\title{A New Approach to Weak Amplitudes in Large-$N_C$ QCD
\thanks{UAB-FT-475, CPT-99/P.3901}}

\author{Marc Knecht\address{Centre de Physique Th{\'e}orique,
        CNRS-Luminy, Case 907, \\
        F-13288 Marseille Cedex 9, France}%
        , 
        Santiago Peris\address{Grup de Fisica Te{\`o}rica and IFAE, 
Universitat Aut{\`o}noma de Barcelona, \\
        E-08193 Bellaterra, Barcelona, Spain} 
and Eduardo de Rafael$^a$}

\begin{document}

\begin{abstract}

We report on some recent progress made in understanding weak matrix 
elements of mesons in the context of the next-to-leading order of the 
large-$N_C$ approximation to QCD. Specifically, we first use the example of 
the weak contributions to the $\pi^+-\pi^0$ mass difference to exhibit how 
a systematic matching can be achieved analytically between short distances 
and long distances within our large-$N_C$ framework. We are then also able to 
compute matrix elements of the operator $Q_7$, as they turn out to
depend on the same QCD correlator as the previous pion mass
difference. As a final example we determine the chiral counterterms 
governing the pseudoscalar decay into a lepton pair, where we briefly 
comment also on the special case $K_L\ra \mu^+ \mu^-$. 
This report covers the material presented by the 
authors in three separate talks at the QCD'99 conference in Montpellier.

\end{abstract}

\maketitle

\section{INTRODUCTION}

At the present level of accuracy, the simple description of electroweak 
interactions in the Standard Model gives an 
excellent fit to the experimental data at high energies collected  over
the last years, for  instance at LEP \cite{treille}. 
For low-energy phenomenology
however, the description in terms of the fields
which appear in the Standard Model Lagrangian is inappropriate.  It is
more convenient to use an effective Lagrangian description where the heavy
degrees of freedom of the Standard Model are integrated out in the presence
of the strong interactions. This procedure leads, however, to a rather
complicated  structure for the weak processes involving hadrons.  The
effective  Hamiltonian that describes the non-leptonic $\Delta S = 1$
decay of kaons, for instance, is given by a set of four-quark
operators $Q_i$, 
\be\lbl{Heff}
\cH^{\eff}\,=\,\sum_i C_i(\mu) \ Q_i(\mu),
\ee
modulated by the functions $C_i(\mu)$, the Wilson coefficients,
containing the information from the short-distance physics. The matching 
of the effective theory to the underlying theory of the Standard Model is 
performed at a high scale 
$\mu_{{\mbox{\tiny SM}}}\sim m_{\mbox{\tiny t}}\sim M_W$, where the 
corresponding Wilson coefficients $C_i(\mu_{{\mbox{\tiny SM}}})$ can be 
accurately computed in perturbation theory. In order to evaluate a weak 
matrix element at much lower energies $\mu\ll\mu_{{\mbox{\tiny SM}}}$, the 
Wilson coefficients are evolved downward using their renormalization 
group properties \cite{buras}.

Physical observables evaluated with the effective Hamiltonian in
Eq.~(\ref{Heff}) are independent of the arbitrary  scale $\mu$  that was
introduced in order to separate the short-distance physics,  contained in
the Wilson coefficients $C_i(\mu)$, from the long-distance  contributions
described by the hadronic matrix elements 
$<h'\vert Q_i(\mu)\vert h>$ of the four-quark operators $Q_i(\mu)$. 
However, depending on the scale $\mu$ at which one is working, the 
contribution of a given operator $Q_i(\mu)$ can be more or less 
important, according to the behaviour of the modulating factor $C_i(\mu)$ 
under the renormalization group evolution. The latter is available only 
within a perturbative expansion in powers of the strong coupling constant 
$\alpha_s$. Although the present state of the art includes NNLO
contributions\cite{buras} ,
it can only provide a reliable description of the scale 
dependence of the Wilson coefficients down to a scale $\mu$ which may, at 
best, be taken slightly below the charm quark mass, $\mu\lesssim 
m_{\mbox{\tiny c}}$. On 
the other hand, the computation of the matrix elements 
$<h'\vert Q_i(\mu)\vert h>$ themselves requires non-perturbative methods.
One possibility is to have recourse to numerical computations based on a 
discretized lattice version of QCD. Another approach consists in implementing 
in an analytic way some particular but systematic 
expansion scheme, like chiral perturbation theory (ChPT) and/or the 
large-$N_C$ expansion, etc. The fact that the final result for the 
matrix element $<h'\vert\cH^{\eff}\vert h>$ must then be independent of the 
factorization scale (at the given order of the non-perturbative expansion 
considered for the evaluation of the matrix elements of the four-quark 
operators) provides thus a non-trivial constraint and a crucial check of 
the whole calculation.
 
Finally, at a very low scale $\mu\ll 1$ GeV, the interactions of the light 
pseudoscalar mesons can be described within a systematic expansion in 
powers of momenta (chiral expansion). Indeed, in the chiral limit, these 
pseudoscalar states are the Goldstone bosons of spontaneous chiral 
symmetry breaking, so that their interaction becomes small at low energy. 
To lowest order in ChPT, their strong and weak matrix elements can be 
obtained, in the chiral limit, from the $\cO(p^2)$ chiral Lagrangian
\be\lbl{Leff}
\cL^{\eff}_{\chi}\,=\,\cL^{(2)}_S + \cL^{(2)}_{W;8} + \cL^{(2)}_{W;27}\,,
\ee\lbl{Ls} 
where $\cL^{(2)}_S$ accounts for the strong interactions between the 
pseudoscalars
\be
\cL^{(2)}_S\,=\,-\frac{1}{F_0^2}\tr(\cL_\mu\cL^\mu),
\ee\lbl{Lw8}
while the weak $\vert\Delta S\vert = 1$ transitions receive contributions with the 
(8,1) quantum numbers of the chiral group $SU(3)_L\times SU(3)_R$
\be
\cL^{(2)}_{W;8}\,=\,-4\frac{G_F}{\sqrt{2}}V_{ud}V_{us}^*g_8
                     (\cL_\mu)_{2i}(\cL_\mu)_{i3}\,+\,{\mbox{h.c.}},
\ee
or with the (27,1) quantum numbers
\bea
\cL^{(2)}_{W;27} &=& -4\frac{G_F}{\sqrt{2}}V_{ud}V_{us}^*g_{27}
\bigg[
                  \frac{2}{3}(\cL_\mu)_{21}(\cL_\mu)_{13}\nn\\
&&\qquad +(\cL_\mu)_{23}(\cL_\mu)_{11}\bigg] \,+\,{\mbox{h.c.}}\,.
\eea
Here we use the notation $\cL_\mu = -i\frac{F_0^2}{2}U^+D_\mu U$, and as 
usual $U$ denotes a unitary $3\times 3$ matrix describing the octet of 
pseudoscalar fields.
All strong and weak matrix elements involving only the Goldstone bosons 
can thus be evaluated at $\cO(p^2)$ accuracy and expressed in terms of 
only three low-energy constants, the decay constant of the pion in the 
chiral limit $F_0$, and the two constants $g_8$ and $g_{27}$ which,
because of CP violation, are complex numbers. 
For instance, in this approximation, we obtain the following  
ratio of the $K\to(\pi\pi)_I$ amplitudes for the decay of kaons into two 
pions in a prescribed isospin channel $I=0,2$
\be\lbl{onehalfrule}
\left\vert\frac{\cA(K\to(\pi\pi)_0)}{\cA(K\to(\pi\pi)_2)}\right\vert\,=\,
\frac{1}{\sqrt{2}}\left\vert\frac{g_8 + g_{27}}{g_{27}}\right\vert .
\ee
Higher order or quark mass corrections can, of course, be considered as well, 
but at the expense of introducing additional low-energy constants.

Unfortunately, an $a~priori$ knowledge of 
the values of the constants $g_8$ and $g_{27}$, which would lead to a 
{\it quantitative} understanding of the ratio in Eq.~ (\ref{onehalfrule}),
i.e.  of the $\Delta I = 1/2$ rule, 
is not available 
at present. Such a knowledge would require that one is actually able to 
continue the process of integrating out the high energy modes in (\ref{Heff}) 
in a {\it non-perturbative way} down to the very low scale where the 
description in terms of the effective Lagrangian (\ref{Leff}) is valid.
On the other hand, these low-energy constants can be identified 
in terms of QCD correlators of bilinear quark operators. For instance, 
$F_0$ is exactly given by the following two-point correlator in the 
chiral limit and at zero momentum transfer:
\be 
\lim_{Q^2\to 0}\left(-Q^2\Pi_{LR}(Q^2)\right)\,=\,F_0^2,
\ee
where
\bea
\lbl{lrtpf}
\Pi_{LR}^{\mu\nu}(q)&=&2i\!\int d^4 x\,e^{iq\cdot x}\langle 0\vert
\mbox{\rm T}\!\left(L^{\mu}(x)R^{\nu}(0)^{\dagger}
\right)\!\vert 0\rangle \nonumber \\
&=& (\!q^{\mu}q^{\nu}\!-\!g^{\mu\nu}q^2)\Pi_{LR}(Q^2)\,,
\eea
with $Q^2 = -q^2$, and where $L^{\mu}$ and $R^{\nu}$ are left- and 
right-handed currents with the appropriate flavour quantum numbers
$e.g.$ $L_\mu\,=\,{\overline{u_L}}\gamma_\mu d_L$,
$R_\mu\,=\,{\overline{u_R}}\gamma_\mu d_R$. The 
low-energy constants of the strong interaction part $\cL^{\eff}_S=\sum_{n\ge 1}
\cL^{(2n)}_S$ describing higher orders in the chiral expansion can 
likewise 
be expressed in terms of the low-momentum behaviour of QCD correlators. 
For instance, one of the Gasser-Leutwyler constants \cite{GL85} 
at order $\cO(p^4)$,
$L_{10}$,  is 
given by the coefficient of the secont term in the expansion of 
$-Q^2\Pi_{LR}(Q^2)$, i.e. 
\be
-Q^2\Pi_{LR}(Q^2)\,=\,F_0^2+4L_{10}Q^2+\cO(Q^4)\,,
\ee 
modulo chiral logarithms that we have not displayed explicitly. 
In general, one can find an analogous representation for the other low-energy 
constants of $\cL^{\eff}_S$ as coefficients of a Taylor expansion around zero
momentum of certain Green functions,
but not always two-point Green functions. Using dispersion 
relations, one may then express these low-energy constants in terms of 
observables of the hadronic spectrum in the chiral limit. 
As a matter of fact, the low-energy 
constants of interest, such as $F_0$, and the corresponding Green's 
functions, such as $\Pi_{LR}^{\mu\nu}(q)$, happen to be {\it order parameters} 
of chiral symmetry breaking.
This means that these Green's functions, 
which receive no contribution from the pertubative QCD continuum, behave 
smoothly at short distances, and that the corresponding 
dispersion relations converge therefore rapidly. In practice, one may 
thus expect to need information on the hadronic spectrum only over a 
finite energy range, say up to $\sim$1 GeV. Unfortunately such data, even 
away from the chiral limit, are not always available. An enormous 
simplification comes about after taking the large-$N_C$ limit of
QCD. In this limit the spectrum consists of an infinite number of 
zero-width mesonic resonances. For instance,
$\Pi_{LR}(Q^2)$ is then given by the pion pole and by an infinite sum over
single particle vector and axial-vector states, i.e.
\bea\label{eight}
-Q^2\Pi_{LR}(Q^2)=F_0^2\qquad\qquad\qquad\qquad \nonumber \\
+Q^2\sum_{A} \frac{f_{A}^2 M_{A}^2}{M_{A}^2+Q^2} - 
Q^2\sum_{V} \frac{f_{V}^2 M_{V}^2}{M_{V}^2+Q^2}\,,
\eea
which, upon expanding the above large-$N_C$ representation of
$-Q^2\Pi_{LR}(Q^2)$ around $Q^2=0$, yields 
the following expression for $L_{10}$ \cite{GL84}\cite{KdeR98}, 
\be\label{L10}
4 L_{10}=\sum_A f_A^2 - \sum_V f_V^2\,,
\ee
in terms of the parameters of the zero-width mesonic resonances.

The situation is more involved when it comes to the constants occurring 
in the $\vert\Delta S\vert = 1$ sector, such as 
$g_8$ or $g_{27}$.
Although they can still be expressed in terms of QCD four-point 
functions of quark bilinears which are order parameters of spontaneous 
chiral symmetry breaking, they involve for instance two left-handed charged 
currents convoluted with a free $W$-boson propagator. 
Therefore, the whole 
range of momenta, and not only the low-momentum region, is involved. The same
happens to the constants describing electromagnetic
corrections to strong processes.


The crucial question is of course whether this interpolation at large $N_C$ 
can be realized in 
a way that provides the correct matching of the scale dependence of the weak 
matric elements $<h'\vert Q_i(\mu)\vert h>$ with the scale dependence of 
the short-distance piece encoded in the Wilson coefficients, $C_i(\mu)$. As we
shall see, there are some observables where we have already been able to
realize this program.

\section{A CLASSIC EXAMPLE REVISITED}

In the chiral limit the $\pi$, the $K$ and the $\eta$ form an octet of
massless Goldstone bosons, provided electroweak interactions are neglected.
In the presence of electromagnetic interactions, however, the electrically 
charged members of the erstwhile 
octet of Goldstone bosons acquire a mass even in the chiral limit. 
This mass term can be described by the effective operator
\bea
\lbl{effem}
\cL^{\eff}_{mass;\gamma} \,=\, 4 \pi \alpha C_\gamma\,\tr
\left(Q_{R}UQ_{L}U^{\dagger}\right)\qquad\qquad \nonumber \\ 
\qquad\,=\, -8 \pi \alpha C_\gamma\frac{1}{F_0^2} (\pi^+
\pi^- + K^+ K^-) + \cdots\,, 
\eea
where $Q_{R,L}$ are matrices of charges in flavour space governing 
the couplings of quarks to the photon, 
$Q_{R}\!=\!Q_{L}\!=\!\mbox{\rm diag}[2/3, -1/3, -1/3]$, and $\alpha$ is the 
fine structure constant. The constant $C_\gamma$ results from
integrating out all degrees of freedom but the Goldstone bosons and the 
photon in the Standard Model Lagrangian. The fact that the QCD part is 
described in terms of quarks and gluons and not in terms of hadrons is 
of course a good part of the difficulty in performing this integration.

In this classic example, it is well known that $C_\gamma$ is
determined by the same 
two-point function $\Pi_{LR}(Q^2)$ that we have already encountered
\cite{Dasetal}, 
but contrary to $F_0$ or to $L_{10}$, 
which describe its low-momentum behaviour,
the identification of $C_\gamma$ involves an {\it integral} 
of $\Pi_{LR}(Q^2)$, 
weighted with the free photon propagator, over the whole range of momenta,
\be
\lbl{Cg}
C_\gamma = -\frac{1}{8\pi^2}\,\frac{3}{4}
\int_0^\infty dQ^2
\,Q^2\Pi_{LR}(Q^2).
\ee
We stress here that $C_{\gamma}$ is thus very
akin to the coupling constants in the chiral Lagrangian describing
electroweak interactions of hadrons, such as $g_8$ or $g_{27}$ for instance.

The integral in (\ref{Cg}) can be 
split-off at an arbitrary intermediate scale $\Lambda$, but large enough that
perturbation theory sets in \cite{bardeenCg}. The high-momentum region 
of integration, $Q^2\gg\Lambda^2$, can be evaluated upon taking for 
$\Pi_{LR}(Q^2)$ its asymptotic behaviour at short distances, which in QCD 
is given by (this property ensures the 
convergence of Eq.~(\ref{Cg}))
\be
\lim_{Q^2\ra\infty} Q^4\Pi_{LR}(Q^2)\,=\,0\,,
\ee
and
\bea\label{six}
\lim_{Q^2\ra\infty} Q^6\Pi_{LR}(Q^2) =\qquad \qquad \qquad \qquad\nonumber\\
-4\pi^2\left(\frac{\alpha_s}{\pi}+ \cO(\alpha_s^2)\right)
\langle\bar{\psi}\psi\rangle^2\,. 
\eea
Notice that in the last formula, we have already replaced the relevant 
four-quark condensate by its factorized large-$N_C$ expression.
For the large-$N_C$ limit representation in Eq.~(\ref{eight}), 
these properties translate into the first and second Weinberg sum
rules \cite{Master}   
\be\lbl{W1}
\sum_{V}f_{V}^2 M_{V}^2-\sum_{A}f_{A}^2 M_{A}^2=F_0^2,
\ee
\be\lbl{W2}
\sum_{V}f_{V}^2 M_{V}^4-\sum_{A}f_{A}^2 M_{A}^4=0,
\ee
and into the relation \cite{KdeR98}
\bea\lbl{asympt}
\sum_{V}f_{V}^2 M_{V}^6-\sum_{A}f_{A}^2 M_{A}^6 = \qquad 
\qquad \qquad \qquad \nonumber \\
-4\pi^2\left(\frac{\alpha_s}{\pi}+ \cO(\alpha_s^2)\right)
\langle\bar{\psi}\psi\rangle^2 \ , \ 
\eea
respectively.
The corresponding contribution to $C_\gamma$ from the high-momentum part 
of the integral reads
\bea
C_\gamma^>(\Lambda)\,=\,+\frac{3}{8}
\left(\frac{\alpha_s}{\pi}+ \cO(\alpha_s^2)\right)
\frac{\langle\bar{\psi}\psi\rangle^2}{\Lambda^2} \nn
\eea
\bea
\qquad \qquad + \cO \left(\frac{1}{\Lambda^4}\right)\,.
\eea
For the low-momentum region of the integral, the large-$N_C$ representation
of Eq.~(\ref{eight}) gives
\bea
C_\gamma^<(\Lambda)\,= \qquad \qquad \qquad \qquad \qquad \qquad \qquad \qquad 
\nonumber \\
+ \frac{3}{32\pi^2}\bigg[\sum_Af_A^2M_A^4\ln\frac{M_A^2}{\Lambda^2}
-\sum_Vf_V^2M_V^4\ln\frac{M_V^2}{\Lambda^2}\bigg]\nonumber \\
+ \frac{3}{32\pi^2} \ \ \frac{\sum_{V}f_{V}^2 M_{V}^6-\sum_{A}f_{A}^2
  M_{A}^6}{\Lambda^2} \qquad \qquad \qquad \nonumber 
\eea
\bea
+ \cO \left(\frac{1}{\Lambda^4}\right)\ , \qquad 
\eea
{\it after} using the two Weinberg sum rules (\ref{W1}) and (\ref{W2}).
Upon adding the two pieces, one obtains a perfect matching to the given order
in $1/\Lambda$, i.e. 
$C_\gamma^<(\Lambda) + C_\gamma^>(\Lambda) \sim \cO(1/\Lambda^4)$, due to 
the sum-rule in Eq.~(\ref{asympt}). 
Clearly this cancellation can be effected to any
order in $1/\Lambda$, as Eq.~(\ref{Cg}) has no $\Lambda$ to begin 
with, but the point
is that, in order to accomplish this $\Lambda$ 
matching, care must be taken that
the constraints imposed by the OPE are satisfied to that order in
$1/\Lambda$ \cite{KdeR98}. 
This lesson is very important in the case of
weak matrix elements. 

Let us notice here that a simple ansatz that realizes all the
properties listed above consists in 
taking into account only one vector and one axial-vector resonance
(satisfying even some other stronger constraints discussed 
in Ref.\cite{PPdeR98}) that we shall call the Lowest Meson Dominance (LMD)
approximation to large-$N_C$ QCD. In this case one has the simpler 
expression \cite{bardeenCg} 
\be\lbl{PiLMD}
-Q^2\Pi_{LR}^{\mbox{\tiny{LMD}}}(Q^2)\,=\,
\frac{M_{\rho}^2 M_{a1}^2 F_0^2}{(Q^2+M_{\rho}^2) (Q^2+M_{a1}^2)}\, ,
\ee
where $M_{\rho}^2=\frac{3\sqrt6}{5N_C}(4\pi F_0)^2$, $M_{a1}=\sqrt{2}
M_{\rho}$ \cite{PPdeR98} and $F_0\simeq 87$MeV is an
estimate of the pion decay constant in the chiral limit 
\cite{GL84}\cite{GL85}. 
Combining Eqs.~(\ref{effem},\ref{Cg}) and (\ref{PiLMD}) one obtains
\bea
m_{\pi^+}^2=F_0^2\ \frac{24\sqrt6}{5}\pi\alpha\ln2 \quad , 
\eea
i.e. $m_{\pi^+}-m_{\pi^0}\simeq\  5$ MeV to be compared 
to $m_{\pi^+}-m_{\pi^0}|_{exp}\simeq 4.6$ MeV.

In order to create a situation which is more similar to the one encountered 
in the case of the $\vert\Delta S\vert = 1$ transitions, let us rather consider the 
masses acquired by the pseudoscalar octet under the influence of the weak 
neutral currents \cite{KPdeR98}. 
They result from an effective term similar to (\ref{effem})
\be
\lbl{effZ}
\cL^{\eff}_{mass;Z^0} = 4 \pi \alpha C_Z\,\tr
\left(Q_{R}UQ_{L}U^{\dagger}\right)\,,
\ee
but with a coupling constant $C_Z$ which involves the integral of 
$\Pi_{LR}(Q^2)$ weighted by the propagator of the massive $Z^0$,
\bea
\lbl{CZ}
C_Z = \frac{1}{8\pi^2}\,\frac{3}{4}
\int_0^\infty dQ^2\,\frac{Q^2}{Q^2+M_Z^2}\,Q^2\Pi_{LR}(Q^2).\nn\\
\eea
The change of sign as compared to Eq. (\ref{Cg}) is a consequence of the
$SU(2)\times U(1)$ quantum numbers of the quarks.
Using an Euclidean momentum cutoff $\Lambda$, one may again 
split the integral into a low-energy piece, which gives
\bea\lbl{low}
C_Z^<(\Lambda)\,=\,\frac{3}{32\pi^2}\frac{1}{M_Z^2}\bigg[
\sum_Af_A^2M_A^6\ln\frac{M_A^2}{\Lambda^2}\nn\\
-
\sum_Vf_V^2M_V^6\ln\frac{M_V^2}{\Lambda^2}\bigg]\,,
\eea
and a high-energy piece, evaluated with the leading short-distance behaviour 
given by Eq.~(\ref{six}),
\be\lbl{high}
C_Z^>(\Lambda)\,=\,-\frac{3}{8\pi}\frac{1}{M_Z^2}\alpha_s
\langle\bar{\psi}\psi\rangle^2\,\,\ln\frac{M_Z^2}{\Lambda^2}.
\ee
Adding Eqs. (\ref{low}) and (\ref{high}) one obtains the final result which
is, of course, $\Lambda$ independent.

A more common approach is to construct an effective Lagrangian in which the
$Z^0$ has been integrated out and the Green functions are renormalized using
the $\overline{MS}$ scheme. The relevant term in the 
Lagrangian of the Standard Model which is
responsible for the $Z$-induced contribution to the
$\pi^{+}-\pi^{0}$ mass difference is the neutral current interaction
term
\bea
\cL_{\NC} =\frac{e}{2\sin\theta_{W}\cos\theta_{W}}
\bigg[\bar{q}_{L}\gamma^{\mu}T_{3}q_{L}\nonumber\\
\qquad
-2\sin^{2}\theta_{W}
\bar{q}_{L}\gamma^{\mu}Q_{L}q_{L}\nonumber\\
\qquad
-2\sin^{2}\theta_{W}
\bar{q}_{R}\gamma^{\mu}Q_{R}q_{R}\bigg]Z_{\mu}\,.
\eea
When looking for the induced effective Lagrangian of order
$\cO(p^0)$ which contributes to Goldstone boson masses, it is
sufficient to consider left-right operators. In the absence of the
strong interactions, the effective four-quark Hamiltonian which
emerges after integrating out the $Z$ field  is
\bea\label{eq:effNC}
-\cH_{\NC}^{\eff} =\frac{-1}{M_{Z}^2}\frac{e^2}{\cos^2 \theta_{W}} 
\times \qquad \qquad \qquad \nn \\
\bigg[\sin^2 \theta_{W} Q_{LR}
-\frac{1}{2}\left(\bar{q}_{L}\gamma_{\mu}T_{3}q_{L}\right)
\left(\bar{q}_{R}\gamma^{\mu}Q_{R}q_{R}\right)\bigg] \nonumber\\
=\frac{e^2}{M_{Z}^2} Q_{LR}\qquad \qquad \qquad \qquad \qquad \qquad \qquad 
\nn \\
-\frac{e^2}{M_{Z}^2}\frac{1}{\cos^2
\theta_{W}}\frac{1}{6}\left(\sum_{q}\bar{q}\gamma_{\mu}q\right) 
\left(\bar{q}_{R}\gamma^{\mu}Q_{R}q_{R}\right)\ , \nn \\
\eea
where
\be
 Q_{LR}\equiv\left(\bar{q}_{L}\gamma_{\mu}Q_{L}q_{L}\right)
\left(\bar{q}_{R}\gamma^{\mu}Q_{R}q_{R}\right)\,,
\ee 
and summation over quark colour indices within brackets is
understood. In fact, to $\cO(p^0)$,
only the first term proportional to the four-quark operator $Q_{LR}$
can contribute. In the presence of the strong interactions, the
evolution of $Q_{LR}$ from the scale
$M_{Z}^2$ down to a scale
$\mu^2$ can be calculated in the usual way, provided this $\mu^2$ is 
still large enough for a perturbative QCD (pQCD) evaluation to be valid. 
In the leading logarithmic approximation in
pQCD, and to leading non--trivial order in the $1/N_C$ expansion, the relevant
mixing in this evolution which we need to retain is simply given by
\bea\label{eq:mixing}
Q_{LR}(M_{Z}^2) = Q_{LR}(\mu^2) \qquad \qquad \qquad \qquad \nn \\
-3\frac{\als}{\pi}\frac{1}{2}\log\frac{M_{Z}^2}{\mu^2}
D_{RL}(\mu^2)\nonumber\\
+\cdots 
\eea
where $\mu$ is the $\overline{MS}$ scale and 
$D_{RL}$ denotes the four-quark density-density operator
\be D_{RL} \equiv
\sum_{q,q'} e_{q}e_{q'}(\bar{q'}_{L}q_{R})(\bar{q}_{R}q'_{L})\,,
\ee with $e_{q}$ and $e_{q'}$ the quark charges in units of the electric
charge. This can be seen as follows: in the $\overline{MS}$ renormalization
scheme,
the full evolution of the   Wilson coefficients $c_{Q}$ and $c_{D}$ of the
operators
$Q_{LR}$ and
$D_{LR}$ at the one-loop level is governed by the equations (subleading contributions in the $1/N_C$ expansion have been neglected)
\bea
\mu^2\frac{d}{d\mu^2}\left( \begin{array}{c} c_{Q} \\ c_{D} \end{array}\right)
=\qquad \qquad \qquad \qquad \qquad \nonumber\\
\frac{1}{4}\frac{\als N_C}{\pi}\left(\begin{array}{cc} \cdots & \cdots \\
\frac{6}{N_C} & -3+\cdots \end{array}\right)\left( \begin{array}{c}
c_{Q} \\ c_{D} \end{array}\right)\,,
\eea
with boundary conditions: $c_{Q}(M_Z)=1$ and
$c_{D}(M_Z)=\frac{3}{2}\frac{\als}{\pi}$. The result in eq.~(\ref{eq:mixing})
follows when taking $c_{D}(M_Z)=0$, 
which is appropriate when keeping the one--loop leading log
only, and from the off--diagonal term in the (transposed) anomalous
dimension matrix.

We are then confronted with a typical problem of bosonization of four-quark
operators.
The bosonization of $D_{RL}$ is only needed to leading order in the
$1/N_C$ expansion. To that order and to order $\cO(p^0)$ in the
chiral expansion it can be readily obtained from the bosonization of
the factorized density currents, with the result~\footnote{See e.g. the
lectures in Ref.~\cite{deR95} and references therein.}
\bea\lbl{bosonizing}
D_{RL} \equiv \sum_{q,q'}
e_{q}e_{q'}(\bar{q'}_{L}q_{R})(\bar{q}_{R}q'_{L})\qquad \ra \nn \\
\left(\frac{\langle{\bar\psi}\psi\rangle}{2}\right)^2\, 
\tr \left(UQ_{L}U^{\dagger}Q_{R}\right)\, .
\eea
We find that the overall contribution of the term proportional to the
$D_{RL}(\mu^2)$ four--quark operator, which we denote 
$C_{Z}\vert_{D_{RL}}$, is
given by the expression
\be\label{eq:4sd}
C_{Z}\vert_{D_{RL}} =\frac{-3}{8\pi}\frac{1}{M_Z^2}
\als \langle\bar{\psi}\psi\rangle^2 \log\frac{M_{Z}^2}{\mu^2}\,,
\ee
and it is exactly the same result as the
one coming from the $d=6$ term of the OPE in the previous calculation of
Eq. (\ref{high}), except for the difference between $\mu$ and $\Lambda$. 

The problem is then reduced to the bosonization of the operator
$Q_{LR}(\mu^2)$. We are confronted here with a typical calculation of a 
hadronic matrix element of a four-quark operator, in this case the matrix
element
$\langle\pi^{+}\vert Q_{LR}(\mu^2)\vert\pi^{+}\rangle$. The factorized 
component
of the operator $Q_{LR}$, which is leading in $1/N_C$, cannot contribute to the
$\cO(p^0)$ term of the low-energy effective Lagrangian. The contribution
we want from this matrix element is therefore the next-to-leading one in
the $1/N_C$ expansion.

The calculation proceeds along much the same lines as first
suggested in papers by Bardeen, Buras and G{\'e}rard~\cite{bardeen}
sometime ago~\footnote{See also  Refs.~\cite{BGK91,FG95,Ko98} and references
therein for more recent work.}, except that we shall go beyond loops
generated by Goldstone particle interactions alone in order to achieve a
correct matching with the logarithmic scale dependence of the
short--distance contribution in eq.~(\ref{eq:4sd}). 
We can evaluate now the matrix element
$\langle\pi^{+}\vert Q_{LR}(\mu^2)\vert\pi^{+}\rangle$
within the framework of an effective Lagrangian which is a
straightforward generalization to an arbitrary number of massive $J^P=1^-$
and $J^P=1^+$ mesonic states of the effective Lagrangian corresponding to
the LMD
approximation to \QCD recently discussed in ref.~\cite{PPdeR98}. Furthermore
we shall do so by using an Euclidean momentum cutoff to study the issue of
regularization. We leave the details of the calculation to our work 
in Ref. \cite{KPdeR98}.  The final result reads
\bea\label{eq:QLRresult}
C_Z|_{Q_{LR}}=
 - \frac{F_0^2}{2 M_Z^2} \langle\pi^{+}\vert
 Q_{LR}(\mu^2)\vert\pi^{+}\rangle  \quad {\mbox{\rm with}} \nn \\
\langle\pi^{+}\vert Q_{LR}(\mu^2)\vert\pi^{+}\rangle=\qquad \qquad \qquad 
 \qquad \qquad 
\nonumber\\
\frac{+1}{16\pi^2}
\bigg\{ \frac{3}{2}\mu^4 + 4\frac{L_{10}}{F_0^2}\mu^6 \qquad \qquad \qquad
\qquad \qquad \nn \\
+\frac{3}{F_0^2}\int_{0}^{\mu^2} dQ^2 Q^6 \qquad \qquad \qquad \qquad
\qquad \nn\\
\quad \qquad \times\bigg[ \sum_{V}\frac{f_{V}^2}{Q^2+M_V^2} -
       \sum_{A}\frac{f_{A}^2}{Q^2+M_A^2}\bigg]\bigg\}\,.
\eea

The contribution of the Goldstone bosons alone corresponds to
the two terms in the first line of the r.h.s. of Eq.~(\ref{eq:QLRresult}). They
display a typical polynomial
dependence with respect to the cut-off $\mu$, which can hardly provide a
reasonably good matching with the logarithmic scale dependence coming from
the short--distance contributions in
$C_{Z}\vert_{D_{RL}}$. In fact, in an $\overline{MS}$ regularization scheme, as
commonly chosen for the evaluation of the short-distance Wilson coefficients,
these power divergences will automatically disappear.  Simply adding higher
resonances does not by itself solve the problem of matching the long and the
short distances either; however, when the information coming from 
the two Weinberg sum rules (\ref{W1},\ref{W2}) and the sum
rule (\ref{L10}) is taken into account, 
the result of Eq.~(\ref{eq:QLRresult}) can indeed be
recast into a form which reproduces Eq.~(\ref{low}). Notice
that in an $\overline{MS}$ regularization scheme, the integral in eq.
(\ref{eq:QLRresult}) should have been understood in $n=4-\epsilon$ 
dimensions and therefore multiplied by $\mu^{4-n}$, $\mu$ being the
$\overline{MS}$ regularization scale; and, of course, all the power
divergences should have been put to zero. The result, when combined with
Eq.~(\ref{eq:4sd}) finally yields the contribution of Eq. (\ref{low}) plus 
Eq. (\ref{high}) as before.
We insist on the fact that, regardless of the regularization one chooses,  
the calculation we have done of the matrix element $\langle\pi^{+}\vert
Q_{LR}(\mu^2)\vert\pi^{+}\rangle$ is an exact calculation to next-to-leading
order in the $1/N_C$ expansion and to $\cO(p^0)$ in the chiral expansion.

\begin{figure}[t]
\centerline{\psfig{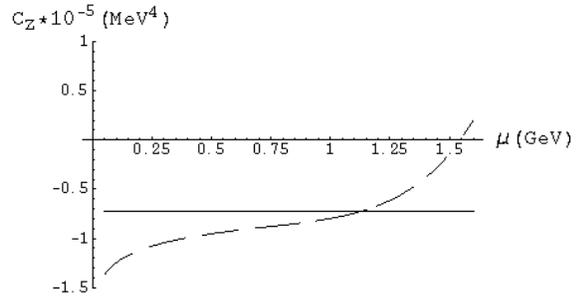}}
\caption[CZplot]
{\small Plot of the total contribution to $C_Z$, Eqs. 
(\ref{eq:4sd}) plus (\ref{eq:QLRresult}), 
i) when only Goldstones are considered
in Eq. (\ref{eq:QLRresult}) (dashed curve), 
ii) full result including all terms in Eq. 
(\ref{eq:QLRresult}) with just one vector and 
axial-vector resonances (solid curve).}
\end{figure}

 In Fig. 1 we compare the complete result (with just 
one resonance for the vector and axial-vector channels)
with the 
value obtained for $C_Z$ when only
 Goldstones are included in the long-distance part, i.e. only the first two
 terms of Eq. (\ref{eq:QLRresult}) are kept. One immediately sees that 
the complete result shows a flat dependence on the cutoff $\mu$, 
as it should, while the result with only Goldstones 
shows a certain dependence on
$\mu$. One also sees that for a
reasonable value of the cutoff $\mu \sim 1.15 $~GeV (which is some sort of
average between the $\rho$ and $a_1$ masses) the result with only Goldstones
gives the right answer but that, if one misses the right value of $\mu$ in
this guess, one can easily even flip the sign of the whole contribution.

We think this clearly 
exemplifies the potential dangers of a calculation with only Goldstones 
and a physical Euclidean cutoff \cite{bardeen}\cite{Ko98}.

\section{ELECTROWEAK PENGUIN OPERATORS}

Within the framework discussed above, we have also
shown~\cite{KPdeR99} that the $K\to\pi\pi$ matrix elements of the
four-quark operator, 
\be\lbl{Q7}
Q_7 = 6(\bar{s}_{L}\gamma^{\mu}d_{L})
\sum_{q=u,d,s} e_{q} (\bar{q}_{R}\gamma_{\mu}q_{R})\,,
\ee
can be calculated to first non--trivial order in the
chiral expansion and in the $1/N_C$ expansion.

The operator $Q_7$ emerges at
the
$M_W$ scale from  considering  the so-called electroweak penguin
diagrams. In the presence of the strong interactions, the renormalization 
group
evolution of
$Q_7$ from the scale $M_W$ down to a scale $\mu\lesssim m_{c}$ mixes this
operator with others, in
particular with the four-quark
density-density operator $Q_8$
\be\lbl{Qeight}
Q_8 = -12\sum_{q=u,d,s}e_{q}(\bar{s}_{L}q_{R})(\bar{q}_{R}d_{L})\,.
\ee
These two operators, times their corresponding
Wilson coefficients, contribute to the lowest order $\cO(p^0)$
effective chiral Lagrangian which induces $\vert\Delta S\vert=1$ transitions
in the presence of electromagnetic interactions to order $\cO(\al)$ and
of virtual
$Z^{0}$ exchange, i.e., the Lagrangian~\cite{BW84}
\bea\lbl{order0}
\cL_{\chi,0}^{\vert\Delta S\vert=1}=-\frac{\GF}{\sqrt{2}}\frac{\al}{\pi}
V_{ud}V_{us}^{\ast}\times\nn \\
\frac{M_{\rho}^{6}}{16\pi^2}\,\h\,\tr
\left(U\lambda_{L}^{(23)}U^{\dagger}Q_{R}\right) +\hc \,,
\eea
where
$\lambda_{L}^{(23)}$ is the effective left--handed flavour matrix
$\left(\lambda_{L}^{(23)}\right)_{ij}=\delta_{i2}\delta_{3j}$
$(i,j=1,2,3)$. This is the only possible invariant which in the
Standard Model can generate
$\vert\Delta
S\vert=1$ transitions to orders
$\cO(\al)$ and $\cO(p^0)$ in the chiral expansion. 
The coupling constant $\h$ is dimensionless and, {\it a priori}, of order
$\cO(N_C^{2})$ in the
$1/N_C$ expansion. It plays a 
crucial r{\^o}le in the phenomenological analysis of radiative corrections
to the
$K\ra\pi\pi$ amplitudes, hence the interest of identifying all the possible
contributions to this constant.

The bosonization of the operator $Q_{7}$ to
next-to-leading order in the $1/N_C$ expansion
turns out to be entirely analogous to the bosonization of the
operator $Q_{LR}\equiv\left(\bar{q}_{L}\gamma^{\mu}Q_{L}q_{L}\right)
\left(\bar{q}_{R}\gamma^{\mu}Q_{R}q_{R}\right)$ which governs the 
electroweak $\pi^{+}\!-\!\pi^{0}$ mass difference  discussed
in the previous section. Because of the $LR$ structure, the factorized 
component of
$Q_{7}$, which is leading in $1/N_C$, cannot contribute to  the 
$\cO(p^0)$  low-energy effective Lagrangian in Eq.~(\ref{order0}). The first
$\cO(p^0)$ contribution from this operator is next-to-leading in the
$1/N_C$ expansion and is given by the integral~\cite{KPdeR99}, 
\bea\lbl{LRgral}
Q_{7}\ra -3ig_{\mu\nu}\int \frac{d^4q}{(2\pi)^4}
\Pi_{LR}^{\mu\nu}(q)\times \nn \\
\tr\left( U\lambda_{L}^{(23)} U^{\dagger} Q_{R}\right)\,,
\eea
involving the {\it same} two-point function as in Eq.~(\ref{lrtpf}).
This integral, however, is divergent for large $Q^2$ and needs to be
regulated. The usual prescription \cite{bardeen} for the evaluation of 
integrals such as this, consists in taking a sharp cut-off in the
(Euclidean)  integration over $Q^2$,
\bea\lbl{mugral}
Q_{7}\ra -6\frac{3}{32\pi^2}\int_{0}^{\Lambda^2} dQ^2
Q^2\left(-Q^{2}\Pi_{LR}(Q^2)\right)\times\nn \\
\!\!\!\!\!\tr\left( U\lambda_{L}^{(23)} U^{\dagger} Q_{R}\right)\,.
\eea
Inserting the same large--$N_C$ expression for the function
$\Pi_{LR}(Q^2)$ as in Eq.~(\ref{eight}), with the short--distance constraints
between the couplings and  masses of the narrow states incorporated, the
integral on the r.h.s. of Eq.~(\ref{mugral}) 
becomes then only logarithmically
dependent on the ultraviolet scale
$\Lambda$, and one obtains  the following result
\bea\lbl{explicit}
\int_{0}^{\Lambda^2} dQ^2 Q^2\left(-Q^{2}\Pi_{LR}(Q^2)\right)= \nn \\
\sum_{A}f_{A}^2 M_{A}^6
\log\frac{\Lambda^2}{M_{A}^2}-\sum_{V}f_{V}^2 M_{V}^6
\log\frac{\Lambda^2}{M_{V}^2}\,, 
\eea
for values of the cutoff much larger than any resonance mass included in the
difference.  
Notice that if only the contribution from the Goldstone pole had been 
taken into account, the resulting expression would have displayed a 
polynomial dependence on the cut-off scale $\Lambda$. 
Another possibility is 
to evaluate the integral in Eq.~(\ref{mugral}) within a dimensional 
regularization scheme, say ${\overline{MS}}$, in which case one obtains the 
same result as in Eq.~(\ref{explicit}), but with the correspondence
between   the cut--off $\Lambda$ and 
the ${\overline{MS}}$ subtraction scale $\mu$ given by   
$\Lambda = \mu\cdot e^{\frac{1}{6}}$. In this case we can check that our
result satisfies the correct renormalization group equation, i.e. by acting
with $\mu d/d\mu$ on $Q_7$ in Eqs. (\ref{mugral},\ref{explicit}) one
obtains  
\bea\lbl{rge}
\mu\frac{d}{d\mu} Q_7 = \frac{-9}{8\pi^2}
\left(\sum_A f_{A}^2 M_{A}^6 - \sum_{V}f_{V}^2 M_{V}^6\right) \nn \\
\times \tr\left( U\lambda_{L}^{(23)} U^{\dagger} Q_{R}\right) \nn \\
=\frac{6}{4\pi} \alpha_s Q_8 \,,\qquad \qquad \qquad  
\eea
where one has to bosonize the operator $Q_8$ in Eq. (\ref{Qeight}) utilizing,
mutatis mutandis, the rule of Eq. (\ref{bosonizing}) and  the constraint of
Eq. (\ref{asympt}).

The bosonic expression of $Q_7$ given by Eqs.~(\ref{mugral}) and
(\ref{explicit}) enables us to compute the $K\to\pi\pi$
matrix elements induced by this operator which, following the usual 
conventions,
we express in terms of the following isospin amplitudes
\be
\langle Q_7\rangle_{I}\equiv \langle (\pi\pi)_{I}\vert Q_7 
\vert K^{0}\rangle\,,
\qquad I=0,2\,.
\ee
To leading order $\cO (p^0)$ in the chiral expansion and to 
next--to--leading
order in the $1/N_C$ expansion, $\cO(1/\sqrt{N_C})$ for $K\ra \pi\pi$
amplitudes, we obtain the result
\bea\label{eq:Q7res}
\lefteqn{\langle Q_7 \rangle_0
= \sqrt{2}\langle Q_7 \rangle_2
=\frac{6\sqrt{3}}{16\pi^2F_0^3}\times} \nn \\ & & 
\!\!\!\!\!\!\!\!\!\!\!\left(\sum_{A}f_{A}^2 M_{A}^6
\log\frac{\Lambda^2}{M_{A}^2}-\sum_{V}f_{V}^2 M_{V}^6
\log\frac{\Lambda^2}{M_{V}^2}\right)\,.
\eea

It has become customary to parameterize the results
of weak matrix elements of four-quark operators
$Q_i$ in terms of the factorized contributions from the
so-called vacuum saturation approximation, modulated by correction
factors, the so called $B$--factors.
Although the resulting $B$ factors for the $\Delta I =1/2$ and
$\Delta I=3/2$ transitions generated by the $Q_7$ operator are found to
depend only logarithmically on the matching scale $\mu$, their actual
numerical values turn out to be  rather sensitive to the precise choice of
$\mu$ in the $\GeV$ region. Furthermore, because of the normalization to
the vacuum saturation approximation inherent to the (rather disgraceful)
conventional definition of $B$-factors, there appears a spurious
dependence on the light quark masses as well. In Fig.~2 we show our
prediction for the ratio
\be\lbl{lattice}
{\widetilde B}_{7}^{(3/2)}\,\equiv\,\frac{\langle\pi^+
\vert Q_7^{(3/2)}\vert K^+\rangle}{\langle\pi^+\vert Q_7^{(3/2)}\vert
K^+\rangle^{\rm VSA}_0}\,,
\ee
versus the matching scale $\mu$ defined in the ${\overline{MS}}$ scheme.
This is also the ratio considered in recent lattice QCD
calculations~\cite{lattice}. [In fact, the lattice definition of
${\widetilde B}_{7}^{(3/2)}$ uses a current algebra relation between the
$K\to\pi\pi$ and the $K\to\pi$ matrix elements which is only valid at
order $\cO (p^0)$ in the chiral expansion.]
In Eq.~(\ref{lattice}), the matrix element in the denominator is evaluated 
in the chiral limit, as indicated by the $0$ subscript .

\begin{figure}[t]
\centerline{\psfig{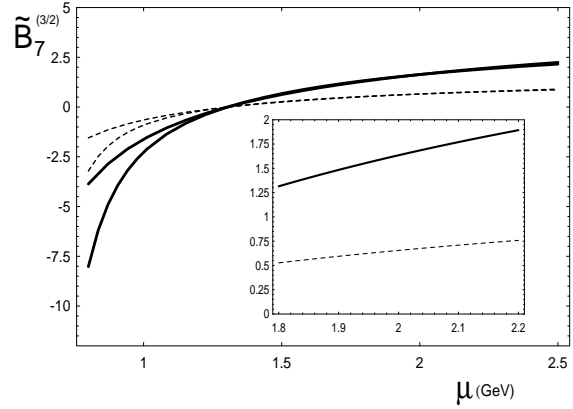}}
\caption[Bplot]
{\small The $\tilde{B}_{7}^{(3/2)}$ factor in
Eq.~(\ref{lattice}) versus $\mu$ in $\GeV$. Solid lines correspond to
$(m_s \!+ \!m_d)(2\,{\rm GeV}) = 158\,{\rm MeV}$; dashed lines
to $(m_s + m_d)(2\,{\rm GeV}) = 100\,{\rm MeV}$.}
\end{figure}

It would be much better, whenever possible, to compare lattice results
directly of the amplitude $\langle\pi^+
\vert Q_7^{(3/2)}\vert K^+\rangle$ with our prediction
\bea\lbl{LMDL}
\lefteqn{\langle\pi^+\vert Q_7^{(3/2)}\vert K^+\rangle  = 
\frac{3}{8\pi^2F_0^2}\!\left[\!\sum_{A}\!\!f_{A}^2
M_{A}^6\log\frac{M_{A}^2}{\Lambda^2+M_{A}^2}
\right. }\nn \\ & & 
\left.-\sum_{V}f_{V}^2 M_{V}^6
\log\frac{M_{V}^2}{\Lambda^2+M_{V}^2}\right]\simeq \nn \\ & & 
\!\!\!\!\!\!\!\! 
\frac{3M_{V}^4}{4\pi^2}\left(\log\frac{\Lambda^2+M_{V}^2}{M_{V}^2}-
2\log\frac{\Lambda^2+2M_{V}^2}{2M_{V}^2}\right)\,, 
\eea
where in the second line we have used the LMD approximation discussed in
Ref.~\cite{PPdeR98}, or the equivalent expression in the
${\overline{MS}}$ regularization, namely, 
\bea\lbl{LMDMSB}
\lefteqn{\langle\pi^+\vert Q_7^{(3/2)}(\mu)\vert K^+\rangle\vert_{\mbox\rm
{\overline{MS}}}  = }\nn \\ & & 
\frac{3M_{V}^4}{4\pi^2}\left(2\log 2-
\log\frac{\mu^2 e^{1/3}}{M_{V}^2}\right)\,.
\eea  
Numerically one obtains $-0.02 \pm 0.01$ GeV$^4$ for the expression in 
Eq. (\ref{LMDMSB}) evaluated at $\mu=2$ GeV. The error 
is an estimate of $1/N_C$ corrections and corrections to the LMD
limit.

It is interesting to compare the $\Lambda$ dependence of Eq. (\ref{LMDL}) with
the $\mu$ dependence of Eq. (\ref{LMDMSB}). 
The two results clearly coincide for
asymptotic values of their respective $\Lambda$ and $\mu$ scales. The
situation, however, is rather different for values of these scales in the
$\GeV$ region. This can be best seen by looking at the functional
relation between these two scales which follows from identifying the two
expressions in Eqs.~(\ref{LMDL}) and (\ref{LMDMSB}): setting
$x\equiv\frac{\mu^2}{M_{V}^2}e^{1/3}$ and
$y\equiv\frac{\Lambda^2}{M_{V}^2}$ this relation is given by the function
$x=4\frac{(1+y/2)^2}{1+y}$. The requirement that $y\ge 0$ results in a
non trivial constraint $x\ge 4$, i.e., $\mu\ge 2
e^{-1/6}M_{V}=1.69M_{V}$. In fact, at the value $\mu=1.69M_{V}$ the
matrix element in Eq.~(\ref{LMDMSB}) flips its sign in contradiction with a 
general positivity property~\cite{Witt83} which demands
that $-Q^2\Pi(Q^2)\ge 0$ for all values of $0\le Q^2\le\infty$. Although
the specific critical value $\mu=1.69M_{V}$ depends on the hadronic LMD
approximation which we have made, it shows that pushing the matching 
of four-quark operators at low
$\mu^2$ values in the  ${\overline{MS}}$ regularization scheme may be very
dangerous and can lead to totally misleading results.

\section{DECAY OF PSEUDOSCALARS INTO LEPTON PAIRS}

The examples discussed so far involved only the two-point function 
$\Pi_{LR}$. In this section we present an example which allows us to 
study the matching between short and long distances in the case where a 
three-point function is involved \cite{Pll}. 
The physical processes of interest are 
the decay of $\pi^0$ or $\eta$ into a pair of charged leptons.  
These processes are dominated by the exchange 
of two virtual photons, as shown in Fig.~3, and it is therefore 
phenomenologically useful to 
consider the branching ratios normalized to the two-photon branching ratio 
($P=\pi^0,\eta$)
\bea
\!\!\!\!\!\!\!\!\!\!
R(P\to\ell^+\ell^-) = \frac{Br (P\to\ell^+\ell^-)}{Br (P\to\gamma\gamma)}
\nonumber\\ 
\qquad
=
2\bigg(\frac{\alpha m_{\ell}}{\pi M_P}\bigg)^2\,
\beta_{\ell}(M_P^2)\,\vert{\cal A}(M_P^2)\vert^2,
\lbl{width}
\eea
with $\beta_{\ell}(s) = \sqrt{1-4m_{\ell}^2/s}$. The unknown dynamics
is then contained in the amplitude ${\cal A}(M_P^2)$.
To lowest order in the chiral expansion the contribution to this
amplitude arises from the two graphs of Fig.~3 with the result      
\bea
\!\!\!\!
{\cal A}(s) = \chi_P(\mu)
+\frac{N_C}{3}\bigg[\,-\frac{5}{2}\nonumber\\
\qquad
+ 
\frac{3}{2}\ln\bigg(\frac{m_{\ell}^2}{\mu^2}\bigg)\,+\,C(s)\,\bigg],
\lbl{amp}
\eea
where 
\be\lbl{counterterm}
\chi_{\pi^0}=\chi_{\eta} = - \frac{(\chi_1+\chi_2)}{4} \equiv \chi \quad  ,
\ee 
with $\chi_1$ and $\chi_2$ the couplings of the two counterterms
which describe the direct  interactions of pseudoscalar mesons with lepton
pairs to lowest order in the chiral expansion~\cite{SLW92}
\bea
&&\!\!\!\!\!\!\!
{\cal L}_{P\ell^+\ell^-}= \frac{3i}{32}\left(\frac{\alpha}{\pi}\right)^2
\,{\bar \ell}\gamma^\mu\gamma_5 \ell
\nonumber\\
&&\!\!\!\!\!\times
\,\big[\chi_1\tr(Q_RQ_RD_\mu UU^{\dagger}
-Q_LQ_LD_\mu U^{\dagger}U)
\nonumber\\
&&\!\!\!\!\!
+\,
\chi_2\tr(U^{\dagger}Q_RD_{\mu}UQ_L-UQ_LD_{\mu}U^{\dagger}Q_R)
\big].\lbl{CTlag}
\eea
The function $C(s)$ in Eq.~(\ref{amp}) corresponds to a finite 
three--point loop integral which
can be expressed in terms of the dilogarithm function 
${\mbox{Li}}_2(x)\,=\,-\int_0^x(dt/t)\ln(1-t)$. 
For $s<0$, its expression reads

\bea
C(s) &=& \frac{1}{\beta_{\ell}(s)}
\bigg[\,{\mbox{Li}}_2\bigg( 
\frac{\beta_{\ell}(s) -1}{\beta_{\ell}(s) +1}
\bigg)\,+
\,\frac{\pi^2}{3}\lbl{Csub}\nonumber\\
&&
+\,\frac{1}{4}\ln^2\bigg(
\frac{\beta_{\ell}(s)-1}{\beta_{\ell}(s)+1}
\bigg)\,\bigg].
\eea

\begin{figure}[t]
\centerline{\psfig{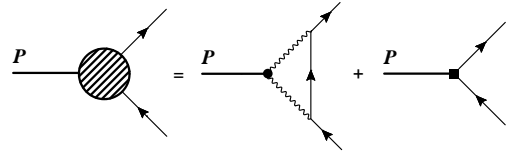}}
\caption[The lowest order contributions]
{\small The lowest order contributions to the 
$P\to\ell^+\ell^-$ decay amplitude. The second graph denotes the 
contribution from the counterterm Lagrangian of Eq.~(\ref{CTlag}).}
\end{figure}

\noindent
The corresponding expression for $s>4m_{\ell}^2$ is obtained by analytic 
continuation, using the usual $i\epsilon$ prescription.
The loop diagram of Fig.~3  
originates 
from the usual coupling of the light pseudoscalar mesons to a photon pair 
given by the well--known Wess--Zumino anomaly~\cite{WZ}. The divergence 
associated with this diagram has been 
renormalized within the ${\overline{\mbox{MS}}}$ minimal subtraction scheme 
of dimensional regularization. The logarithmic dependence on the 
renormalization scale $\mu$ 
displayed in the above expression is compensated by the scale dependence of 
the combination $\chi(\mu)$ of renormalized low-energy constants defined 
above.
Let us stress here that, as shown explicitly in 
Eq.~(\ref{amp}) and in contrast 
with the usual situation in the purely mesonic sector, this scale dependence 
is not suppressed in the large--$N_C$ limit, since it does not arise from 
meson loops.

As a first step towards its subsequent evaluation
we shall identify the coupling constant $\chi$ in 
terms of a QCD correlation function. For that purpose, consider the matrix 
element of the light quark isovector 
pseudoscalar density 
$P^3(x)\,=\,{1\over 2}({\bar u}i\gamma_5 u - {\bar d}i\gamma_5 d)(x)$ 
between leptonic states in the chiral limit.
In the absence of weak interactions, and to lowest non-trivial 
order in the fine structure constant, this matrix element is given by
the integral
\bea
&&\!\!\!\!\!\!\!\!
<\ell^-(p')\,\vert\,P^3(0)\,\vert\,\ell^-(p)>\qquad\nonumber\\
&&
=e^4\int \frac{d^4q}{(2\pi)^4}\,
\frac{{\bar u}(p')\gamma^\mu[{\not\! p}'- {\not\! q} + m_{\ell}]
\gamma^\nu u(p)}{[(p'-q)^2-m_{\ell}^2]q^2(p'-p-q)^2}\nonumber
\\
&&\quad\times
\,i\int d^4x\int d^4y\,e^{iq\cdot x}e^{i(p'-p-q)\cdot y}\nonumber\\
&&\quad\times
<0\,\vert\,T\{j_\mu^{\mbox{\scriptsize{em}}}(x)
              j_\nu^{\mbox{\scriptsize{em}}}(y)P^3(0)\}\,\vert\,0>,
\lbl{matel}
\eea
with 
$j_\mu^{\mbox{\scriptsize{em}}}={1\over 3}
(2{\bar u}\gamma_\mu u - {\bar d}\gamma_\mu d - {\bar s}\gamma_\mu s)$.
In the chiral limit, the QCD three--point correlator appearing in this 
expression is again an order parameter of spontaneous chiral symmetry 
breaking. 
Bose symmetry and parity conservation of the strong interactions yield
\bea\lbl{threepoint}
&&\!\!\!\!\!\!\!\!\!\!
\int d^4x\int d^4y\,e^{iq_1\cdot x}e^{iq_2\cdot y}
\nonumber\\
&&\times
<0\,\vert\,T\{j_\mu^{\mbox{\scriptsize{em}}}(x)
              j_\nu^{\mbox{\scriptsize{em}}}(y)P^3(0)\}\,\vert\,0>\nonumber\\
&&
=\ \frac{2}{3}\,\epsilon_{\mu\nu\alpha\beta}q_1^\alpha q_2^\beta
\,{\cal H}(q_1^2,q_2^2,(q_1+q_2)^2),\lbl{VVP}
\eea
with 
${\cal H}(q_1^2,q_2^2,(q_1+q_2)^2) = {\cal H}(q_2^2,q_1^2,(q_1+q_2)^2)$. 
For very large (Euclidian) 
momenta, the leading short-distance behaviour of this correlation function 
is given by
\bea\lbl{opeone}
&&\!\!\!\!
\lim_{\lambda\to\infty}\,{\cal H}
\big((\lambda q_1)^2,(\lambda q_2)^2,(\lambda q_1 + \lambda q_2)^2\big)
\nonumber\\
&&
=
\ -\frac{1}{2\lambda^4}<{\bar \psi}\psi>\,
\frac{q_1^2+q_2^2+(q_1+q_2)^2}{q_1^2q_2^2(q_1+q_2)^2}\nonumber\\
&&
\, +
\,{\cal O}\bigg(\frac{\alpha_s}{\lambda^4},\frac{1}{\lambda^6}\bigg).
\eea
Actually, what matters for the convergence of the integral in 
Eq.~(\ref{matel}) 
is the leading short-distance singularity of the $T-$product of the two 
electromagnetic currents, which corresponds to
\bea\lbl{opetwo}
&&\!\!\!\!\!\!\!\!
\lim_{\lambda\to\infty}\,{\cal H}
\big((\lambda q)^2,(p'-p-\lambda q)^2,(p'-p)^2\big)
\nonumber\\
&&
=
\ -\frac{1}{\lambda^2}<{\bar \psi}\psi>\,
\frac{1}{q^2(p'-p)^2}
\nonumber\\
&&
 +\,
{\cal O}\bigg(\frac{\alpha_s}{\lambda^2},\frac{1}{\lambda^3}\bigg),
\lbl{high2}
\eea
Thus, the loop integral in Eq.~(\ref{matel}) is indeed 
convergent. The QCD corrections
of order ${\cal O}(\alpha_s)$ in Eqs.~(\ref{high}) and 
(\ref{high2}) will not be considered here. 
Let us 
however notice that since the pseudoscalar density $P^3(x)$ and the 
single-flavour $<{\bar \psi}\psi>$ condensate 
share the same anomalous dimension, the power-like 
fall-off displayed by Eqs.~(\ref{high}) and 
(\ref{high2}) is canonical, i.e. it is 
not modified by powers of logarithms of the momenta.

At very low momentum transfers, the same correlator can be 
computed within 
Chiral Perturbation Theory (ChPT). At lowest order, it is saturated by 
the pion-pole contribution, given by the anomalous coupling of a neutral 
pion, emitted by the pseudoscalar source $P^3(0)$, 
to the two electromagnetic currents, i.e.
\bea
{\cal H}(0,0,(q_1+q_2)^2) &=& \,\frac{N_C}{8\pi^2}\,\frac{<{\bar \psi}\psi>}
{F_0^2}\,\frac{1}{(q_1+q_2)^2}\nonumber\\
&& +\cdots\ ,\lbl{chiral}
\eea
where the ellipsis stands for higher orders in the low-momentum expansion. 
The matrix element $<\ell^-(p')\,\vert\,P^3(0)\,\vert\,\ell^-(p)>$ itself 
may also be evaluated in ChPT. At lowest order, it is given by the diagrams 
of Fig.~3, where the (off-shell) pion is now emitted by the pseudoscalar 
source $P^3(0)$. The result reads, with $t=(p'-p)^2$,
\bea
&&\!\!\!\!\!\!\!\!
<\ell^-(p')\,\vert\,P^3(0)\,\vert\,\ell^-(p)>
\big\vert_{\mbox{~\scriptsize{ChPT}}}\\
&&
=\  - \frac{ie^4}{32\pi^4t}\,
\frac{<{\bar \psi}\psi>}{F_0^2}\,m_{\ell}{\bar u}(p')\gamma_5u(p)
\,{\cal A}(t),\nonumber
\eea
with the function ${\cal A}(t)$ defined in Eqs.~(\ref{amp}) and 
(\ref{Csub}).
The contribution of the loop diagram of Fig.~3 is obtained upon replacing,
in Eq.~(\ref{matel}), the three-point QCD correlator by its 
lowest order chiral expression given in Eq.~(\ref{chiral}).
The coupling constant $\chi(\mu)$ is thus 
given by the residue of the pole at $t=0$ of the matrix element 
$<\ell^-(p')\,\vert\,P^3(0)\,\vert\,\ell^-(p)>$, after subtraction of the 
contribution of the two-photon loop, i.e.
\bea
\frac{\chi(\mu)}{32\pi^4}
\,\frac{<{\bar \psi}\psi>}{F_0^2}\,m_{\ell}{\bar u}(p')\gamma_5 u(p) \qquad
\qquad \qquad \qquad  
\nonumber\\
=\ -\frac{2i}{3}\,{\bar u}(p')\gamma_{\lambda}\gamma_5 u(p) \qquad \times 
\qquad
\qquad \qquad  
\nonumber\\
\lim_{(p'-p)^2\to 0}\int \frac{d^dq}{(2\pi)^d}
\frac{(p'-p)^2}{[(p'-q)^2-m_{\ell}^2]q^2(p'-p-q)^2}\nonumber\\
\times\,
(p'-p-2q)_{\alpha}\bigg[\,q^{\alpha}(p'-p)^{\lambda}
-\,(p'-p)^{\alpha}q^{\lambda}
\,\bigg]\nonumber
\eea
\bea\lbl{exact}
\!\!\!\!
\times\,
\bigg[\,{\cal H}(q^2,(p'-p-q)^2,(p'-p)^2)\nonumber\\
\!\!\!\!
\quad-\,{\cal H}(0,0,(p'-p)^2)\,\bigg].\quad 
\eea
In the large-$N_C$ limit, the three--point correlator (\ref{threepoint}) 
is given by the tree--level 
exchanges of vector and pseudoscalar resonances, as shown in Fig.~4, 
so that the  singularities in each channel consist of a succession of
 simple poles.
This involves couplings of the resonances 
among themselves and to the external sources which, just like the masses 
of the resonances themselves, cannot be fixed in the absence of an
explicit solution of QCD in the large-$N_C$ limit. 
Nevertheless, the general 
structure of the quantity appearing in Eq.~(\ref{exact}) is of the form
\bea\lbl{largeN}
\lim_{(p'- p)^2\to 0}\,
(p'-p)^2\,{\cal H}(q^2,q^2,(p'-p)^2)= \nonumber\\
\ -\frac{1}{2}\frac{<{\bar \psi}\psi>}{F_0^2} \qquad \qquad  
\qquad \qquad 
\nonumber\\ 
\times \ \sum_V\,M_V^2\,\bigg[\,\frac{a_V}{(q^2-M_V^2)}\,-
\,\frac{b_V q^2}{(q^2-M_V^2)^2}\,\bigg] \ ,
\eea
where {\it a priori} the sum extends over the {\it infinite} 
spectrum of vector 
resonances of QCD in the large-$N_C$ limit. Equation (\ref{largeN}) follows 
from 
the fact that its left-hand side enjoys some additional properties:~i)~In 
the pseudoscalar channel, only the pion pole 
survives, while massive pseudoscalar resonances cannot contribute. ii)~The 
momentum transfer in the two vector channels is the same. iii)~Its 
high--energy behaviour is fixed by Eq.~(\ref{high2}).

\begin{figure}[t]
\centerline{\psfig{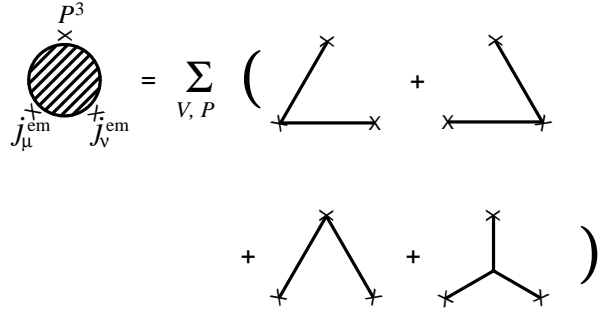}}
\caption[Large-Nc]
{\small 
The con\-tri\-bu\-tions to the ve\-ctor--\-vec\-tor--\-pseu\-do\-sca\-lar 
three--\-point func\-tion in the 
large--\-$N_C$ limit of QCD. The sum extends 
over the infinite number of zero-width vector ($V$) and pseudoscalar ($P$) 
states.
}
\end{figure}

\noindent
Even though the constants $a_V$ and $b_V$ 
depend on the masses and couplings of the vector resonances in an unknown 
manner, they are however constrained by the two conditions 
\be
\sum_V\,a_V = \frac{N_C}{4\pi^2}
, \quad 
\sum_V\,(a_V\,-\,b_V)\,M_V^2 = 2F_0^2,\lbl{cond}
\ee
which follow from Eqs.~(\ref{chiral}) and 
(\ref{high2}), respectively. Notice that 
there are no contributions from the perturbative QCD continuum to these sums.
Taking the first of these conditions 
(which, coming from the anomaly, has no ${\cal O}(\alpha_s)$ corrections) 
into account, we obtain
\be
\chi(\mu) = \frac{5N_C}{12}
-2\pi^2\,\sum_V\,
\bigg[\,a_V\ln\left(\frac{M_V^2}{\mu^2}\right)\,-\,b_V\,\bigg].
\lbl{mainAppr}
\ee
This equation, together with the two conditions (\ref{cond}), constitutes 
the central result of Ref. \cite{Pll}. This is as far as the large-$N_C$ limit 
allows us to go. Let us point out 
that the scale dependence of $\chi(\mu)$ is correctly 
reproduced by the expression (\ref{mainAppr}), 
again as a consequence of the 
first condition in Eq.~(\ref{cond}).

Within the LMD approximation of large--$N_C$ QCD, it is easy to 
write down the expression of the correlation function 
${\cal H}(q_1^2,q_2^2,(q_1+q_2)^2)$ which 
correctly interpolates between the high energy behaviour in 
Eq.~(\ref{high}) 
and the 
ChPT result in Eq.~(\ref{chiral})~\cite{fnte2}
\bea
&&
{\cal H}^{\mbox{\scriptsize LMD}}(q_1^2,q_2^2,(q_1+q_2)^2)\ =\ 
-\frac{1}{2}<{\bar \psi}\psi>\nonumber\\
&&\times
\frac{q_1^2+q_2^2+(q_1+q_2)^2-M_V^4a_V^{\mbox{\scriptsize LMD}}/F_0^2}
{(q_1^2-M_V^2)(q_2^2-M_V^2)(q_1+q_2)^2}.
\lbl{HvLMD}
\eea
Notice that this expression also correctly reproduces the behaviour in 
Eq.~(\ref{high2}).
In this approximation, the two conditions (\ref{cond}) 
completely pin down the 
two quantities $a_V$ and $b_V$ in terms of $F_0$ and of the mass $M_V$ of 
this lowest lying vector meson octet,
\be
a_V^{\mbox{\scriptsize LMD}}=\frac{N_C}{4\pi^2}\quad\mbox{and}
\quad b_V^{\mbox{\scriptsize LMD}}=\frac{N_C}{4\pi^2}\,-
\,\frac{2F_0^2}{M_V^2}.\lbl{abLMD}
\ee

With the results of Eq.~(\ref{abLMD}), and for $N_C$=3, it follows from 
Eq.~(\ref{mainAppr}) that
\be
\chi^{\mbox{\scriptsize LMD}}(\mu) =
\frac{11}{4}\,-\,\frac{3}{2}
\ln\left(\frac{M_V^2}{\mu^2}\right)
-4\pi^2\frac{F_0^2}{M_V^2}.
\lbl{chiLMD}
\ee
Numerically, using the physical values $F_0\simeq 87$~MeV and 
$M_V=M_{\rho}=770$~MeV, we obtain
\be
\chi^{\mbox{\scriptsize LMD}}(\mu = M_V)
\ =\ 2.2\pm 0.9,\lbl{numLMD}
\ee
where we have allowed for a systematic theoretical error of 40\%, as a rule 
of thumb 
estimate of the uncertainties attached to the large-$N_C$ and LMD 
approximations.
The predicted ratios of branching ratios in
Eq.~(\ref{width}) which follow from this result \cite{fnte1} 
are displayed in Table 1. We conclude that, 
within errors, the LMD--approxi\-ma\-tion to 
large-$N_C$ QCD
re\-pro\-duces well the observed rates of pseudoscalar mesons decaying into 
lepton pairs.

\begin{table}[b]
\caption[Results]
{\small The values for the ratios $R(P\to\ell^+\ell^-)$ obtained 
within the LMD approximation to large--$N_C$ QCD and the comparison with 
available experimental results.}
{\small
\begin{tabular}{ccc}
\hline\hline
$R$ & LMD & Experiment \\ \hline
$R(\pi^0\to e^+e^-)\times 10^{8}$ & 
$6.2\pm 0.3$ & $7.13\pm 0.55$~\cite{exp1}
  \\ 
$R(\eta\to \mu^+\mu^-)\times 10^{5}$ & $1.4\pm 0.2$ & $1.48\pm 0.22
$~\cite{exp2} \\ 
$R(\eta\to e^+e^-)\times 10^{8}$ & $1.15\pm 0.05$ & ---\\
\hline\hline 
\end{tabular}
}
\label{table1}
\end{table}

At first sight this approximation may resemble good old Vector
Meson Dominance (VMD). However, there are important differences. Firstly, 
the systematic use of the $1/N_C$
expansion in QCD justifies the use of single particle intermediate 
states (and substantiates what otherwise is only an ansatz in VMD).  
And secondly, the
use of the OPE resolves certain difficulties in the traditional VMD
phenomenological approach, such as e.g. the ambiguity in the use of 
the VMD prescription for just one photon  
or for both photons in the decay $P\ra \ell^+ \ell^-$. 
This ambiguity is important, 
since making the wrong choice 
may result in a violation of the OPE constraints given by
Eqs. (\ref{opeone},\ref{opetwo})   \cite{Ametller}.

The situation in the case of the $\vert\Delta S\vert = 1$ decay 
$K_L^0\to\ell^+\ell^-$ is slightly 
more delicate. It has recently been shown~\cite{GDP98} that  
these processes
can also be described by the expressions (\ref{width}) and 
(\ref{amp}), 
but with an effective constant 
$\chi_{K^0_L}$ containing an additional piece
from the short-distance contributions~\cite{BF97}. Of course, a cast-iron
understanding of these transitions is very important \cite{K0L} as  
the evaluation of
$\chi(\mu)$ could then have a potential impact on possible constraints
on physics beyond the Standard Model. 
At present, the most accurate experimental determination of
the $K_L^0\to\mu^+\mu^-$ branching ratio~\cite{BNL99} gives the
result: 
$Br(K_L^0\to\mu^+\mu^-)=(7.18\pm 0.17)\times 10^{-9}$.
Using the experimental branching ratio~\cite{exp2} 
$Br(K_L^0\to\gamma\gamma)=(5.92\pm 0.15)\times 10^{-4}$, this 
leads to a unique solution for an
{\it effective} $\chi_{K_L^0}=5.17\pm 1.13$. The authors of Ref. \cite{GDP98}
argue that, to a good approximation, one has 
\be\lbl{chiK}
\chi_{K_L^0}=\chi - \epsilon_{\gamma\gamma}\ \delta \chi_{SD}\,,
\ee
where $\chi$ is the constant defined in Eq. (\ref{counterterm}) and 
$\epsilon_{\gamma\gamma}$ is the {\it sign} of the on-shell 
form factor $c(0,0)$ 
(for the notation, we refer to \cite{GDP98}; see in particular Eqs. (6) 
and (7) of that reference) describing the $K_L^0\to\gamma\gamma$ decay. 
However, as also discussed 
in Ref. \cite{GDP98}, the decay $K_L\ra \gamma \gamma$
itself is rather problematic. 
The problem comes from the fact that at lowest order in the 
$SU(3)_L\times SU(3)_R$ chiral expansion, one obtains $c(0,0)=0$.
There are two ways to bypass this situation, 
either by considering 
higher orders, 
or by including the $\eta '$ as an explicit degree of freedom 
already at leading order, 
which can be done within the framework of the 
$U(3)_L\times U(3)_R$ combined chiral and large-$N_C$ expansions along the
lines suggested in Refs. \cite{central}. 
In both cases, additional contributions 
to the 
counterterm Lagrangian in (\ref{CTlag}), and thus to
the effective constant $\chi_{K_L^0}$, have to be considered: 
quark mass corrections in the first case, $1/N_C$ corrections 
in the second. To the best of our knowledge, an 
accurate estimate of either of these 
corrections has, unfortunately, not been attempted so far. 
The analysis in the $U(3)_L\times U(3)_R$ framework 
performed in Ref.~\cite{GDP98} obtains the value of $c(0,0)$ from the
experimental number for $K_L\rightarrow \gamma \gamma$ 
and determines its sign, on combined large-$N_C$ and 
phenomenological grounds, to be
positive. In a large-$N_C$ calculation this experimental value for $c(0,0)$ 
would require going beyond the leading term because in the strict limit
$N_C\rightarrow \infty$ --- keeping $m_u=m_d=0, m_s\not= 0$ 
for simplicity---  
one finds again $c(0,0)=0$. Consistency
demands, then, that subleading terms be also included in the counterterms 
 which contribute to $\chi_{K_L^0}$, 
something that goes beyond the scope of this work \cite{VenezianoShore}.  
If, on the other hand, one accepts the plausible phenomenological arguments of
Ref. \cite{GDP98} whereby these subleading terms are neglected, our result 
(\ref{numLMD})
leads then to $\chi_{K_L^0}=0.4\pm 1.1$, corresponding to a ratio 
$R(K_L^0\to\mu^+ \mu^-)=(2.24\pm 0.41)\times 10^{-5}$ which is $2.5\sigma$ 
above the experimental value 
$R(K_L^0\to\mu^+ \mu^-)=(1.21\pm 0.04)\times 10^{-5}$.

\vskip 0.5cm

{\bf Acknowledgements}

We thank Ll. Ametller, J. Bijnens, V. Gim{\'e}nez, J.I. Latorre, L. Lellouch, 
A. Pich and J. Prades for discussions and M. Perrottet for a very pleasant 
collaboration.  

Work supported in part by TMR, EC-Contract No. ERBFMRX-CT980169 
(EURODA$\phi$NE). The work of S.P. has also been partially supported 
by research project CICYT-AEN98-1093.


\begin{thebibliography}{9}
\bibitem{fnte2}
         A similar analysis for the 
         vec\-tor--\-vec\-tor--\-sca\-lar and 
         vec\-tor--\-axial--\-pseudo\-sca\-lar three--\-point 
         functions can be found in Refs.~\cite{MS94} and \cite{M97}, 
         respectively.
\bibitem{fnte1}
        In the case of the 
        decay into a muon pair, $m_{\mu}/\Lambda_H$ 
        corrections lower the result of 
        Eq.~(\ref{numLMD}) by about 7\%, and are taken into account in the 
        numerical values of Table I.
\bibitem{K0L}
         We refer the reader to Ref.~\cite{GDP98} for further details. See 
         also Refs. \cite{dAmb98} and \cite{Val98} for an alternative 
         approach.
\end{thebibliography}
\end{document}